\documentclass[aps,twocolumn,superscriptaddress,showpacs]{revtex4-1}

\usepackage{graphicx}
\usepackage{dcolumn}
\usepackage{bm}
\usepackage{amsmath}
\usepackage{color}
\usepackage{ulem}
\usepackage{threeparttable}
\usepackage{enumerate}
\usepackage{paralist}

\begin{document}

\title{Low-temperature transport properties of intermetallic compound HoAgGe with kagome spin ice state}

\author{N. Li}
\affiliation{Hefei National Research Center for Physical Sciences at the Microscale, University of Science and Technology of China, Hefei, Anhui 230026, People's Republic of China}

\author{Q. Huang}
\affiliation{Department of Physics and Astronomy, University of Tennessee, Knoxville, Tennessee 37996-1200, USA}

\author{X. Y. Yue}
\affiliation{Institute of Physical Science and Information Technology, Anhui University, Hefei, Anhui 230601, People's Republic of China}

\author{S. K. Guang}
\affiliation{Department of Physics and Key Laboratory of Strongly-Coupled Quantum Matter Physics (CAS), University of Science and Technology of China, Hefei, Anhui 230026, People's Republic of China}

\author{K. Xia}
\affiliation{Department of Physics and Key Laboratory of Strongly-Coupled Quantum Matter Physics (CAS), University of Science and Technology of China, Hefei, Anhui 230026, People's Republic of China}

\author{Y. Y. Wang}
\affiliation{Institute of Physical Science and Information Technology, Anhui University, Hefei, Anhui 230601, People's Republic of China}

\author{Q. J. Li}
\affiliation{School of Physics and Material Sciences, Institute of Physical Science and Information Technology, Anhui University, Hefei, Anhui 230601, People's Republic of China}

\author{X. Zhao}
\affiliation{School of Physical Sciences, University of Science and Technology of China, Hefei, Anhui 230026, People's Republic of China}

\author{H. D. Zhou}
\email{hzhou10@utk.edu}
\affiliation{Department of Physics and Astronomy, University of Tennessee, Knoxville, Tennessee 37996-1200, USA}

\author{X. F. Sun}
\email{xfsun@ustc.edu.cn}
\affiliation{Department of Physics and Key Laboratory of Strongly-Coupled Quantum Matter Physics (CAS), University of Science and Technology of China, Hefei, Anhui 230026, People's Republic of China}
\affiliation{Institute of Physical Science and Information Technology, Anhui University, Hefei, Anhui 230601, People's Republic of China}
\affiliation{Collaborative Innovation Center of Advanced Microstructures, Nanjing University, Nanjing, Jiangsu 210093, People's Republic of China}

\date{\today}

\begin{abstract}

We study the magnetic susceptibility, magnetization, resistivity and thermal conductivity of intermetallic HoAgGe single crystals at low temperatures and in magnetic fields along the $a$ and $c$ axis, while the electric and heat currents are along the $c$ axis. The magnetization curves show a series of metamagnetic transitions and small hysteresis at low field for $B \parallel a$, and a weak metamagnetic transition for $B \parallel c$, respectively. Both the magnetic susceptibility and $\rho(T)$ curve show anomalies at the antiferromagnetic transition ($T\rm_N \sim$ 11.3 K) and spin reorientation transition ($\sim$ 7 K). In zero field and at very low temperatures, the electrons are found to be the main heat carriers. For $B \parallel a$, the $\rho(B)$ curves display large and positive transverse magnetoresistance (MR) with extraordinary field dependence between $B^2$ and $B$-linear, accompanied with anomalies at the metamagnetic transitions and low-field hysteresis; meanwhile, the $\kappa(B)$ mainly decrease with increasing field and display some anomalies at the metamagnetic transitions. For $B \parallel c$, there is weak and negative longitudinal MR while the $\kappa(B)$ show rather strong field dependence, indicating the role of phonon heat transport.

\end{abstract}


\maketitle

\section{Introduction}

It is well known that the kagome spin ice has been realized in artificial spin ice based on two-dimensional arrays of single-domain ferromagnetic islands \cite{NatPhys832, Nature439, PhysRevB094418, Natphys359, NatPhys68, Nature553, RevModPhys1473, NatNanotechnol53}. It is a challenge to explore the abundant phase diagram of spin ice in the thermodynamic limit due to the large magnetic energy scales and the small system sizes \cite{RevModPhys1473, NatNanotechnol53}. Alternatively, the pyrochlore titanites such as Dy$_2$Ti$_2$O$_7$ and Ho$_2$Ti$_2$O$_7$ have been found to exhibit the kagome ice state with a magnetic field applied along the [111] direction \cite{JPhysCondensMatter14, PhysRevLett257205, NatPhys566, PhysRevB.87.144404}. In such kagome ice state, the magnetic field pins the spins on the triangular layers and the ice rule can still be satisfied for the in-plane components of the spins on the kagome layers in a narrow range of magnetic field, which was ascribed to the weak exchange or dipolar interactions in systems. In contrast, the intermetallic compound HoAgGe, which exhibits a ground state of antiferromagnetic (AF) order and displays a sequence of field-induced magnetization plateaus, was found to have a naturally existing kagome spin ice at low temperatures \cite{zhao2020realization}.

HoAgGe belongs to the $R$AgGe ($R$ = Tb -- Lu) series with the ZrNiAl-type structure, an ordered variant of the hexagonal Fe$_2$P family ($P$-62$m$ space group), which is non-centrosymmetric with the rare-earth ions forming a two-dimensional and distorted kagome lattice of corner-sharing equilateral triangles (along the $ab$ plane) \cite{Gibson1996, Baran1998, Morosan2004RAgGe}. The special spin structure leads to the peculiar metamagnetic transitions in these compounds and has attracted much research attentions on both experimental studies and theoretical analysis. The earlier studies suggested that in these compounds the spin anisotropy, induced by the crystal-electric field (CEF), changes from axial (in TbAgGe) to extreme planar (in TmAgGe), which provides the opportunity of investigating the angular dependence of metamagnetism and studying how the phase diagrams vary with the anisotropy \cite{PhysRevB.71.014445, Morosan2004RAgGe}. In this family, HoAgGe is a special one having a natural kagome spin ice state at low temperatures, which is constructed by combining the single-ion axial anisotropy in the hexagonal plane and the effective nearest-neighbour ferromagnetic exchange interactions. The spin ice state evolves into ordered and partially disordered magnetic states with applying magnetic field, which all obeys the kagome ice rule requiring a local ``two-in, one-out" or ``one-in, two-out" spin configuration in each triangular unit \cite{zhao2020realization}. It was found that the magnetic structures at the magnetization plateaus can be obtained from the ground state by reversing certain Ho$^{3+}$ spins with the Ising-like anisotropy, and always satisfy the ice rule. Namely, the metamagnetic transitions of HoAgGe originate from the competition between the external magnetic field and the weaker further-neighbor coupling that does not break the kagome ice rule \cite{zhao2020realization}. The metallic characteristics of HoAgGe may lead to some other exotic phenomena such as the interaction between electrons and the magnetic monopoles expected for the kagome ice. So it is expected that the low-temperature transport properties of this system would exhibit some interesting phenomena associated with its peculiar magnetic properties.

Low-temperature heat transport has been found to be a useful tool to probe the field-induced magnetic transitions, due to either the changes of magnetic excitation transport or the spin-phonon coupling \cite{PhysRevB.63.214407, PhysRevLett.98.107201, PhysRevLett.100.137202, PhysRevLett.102.167202, PhysRevB.82.094405, PhysRevB.83.174518, PhysRevB.85.134412, PhysRevB.86.174413, PhysRevB.95.224419, PhysRevB.99.104419, PhysRevB.104.104403}. In this paper, we carried out the electrical resistivity and thermal conductivity (along the $c$ axis) measurements on the HoAgGe single crystal at low temperatures and in high magnetic fields. The magnetization curves display a series of field-induced phase transitions at low temperature for $B \parallel a$, in contrast to the case of $B \parallel c$, which reveals one weak metamagnetic transition. For $B \parallel a$, large and positive transverse magnetoresistance (MR) with extraordinary field dependence is observed and the $\rho(B)$ curves exhibit anomalies at the metamagnetic transitions, accompanied with low-field hysteresis; meanwhile, the $\kappa(B)$ mainly decrease with increasing field and display some anomalies at the metamagnetic transitions. For $B \parallel c$, weak and negative longitudinal MR is observed while the $\kappa(B)$ show rather strong field dependence indicating the role of phonon heat transport.

\section{Experiments}

HoAgGe single crystals were grown by a flux method. The as-grown crystals are mostly rod-shaped with the longest dimension along the $c$ axis. DC magnetic susceptibility ($\chi$) was measured using a Quantum Design SQUID-VSM. Magnetization with magnetic field up to 14 T was measured using a physical properties measurement system (PPMS, Quantum Design). A rectangular parallelepiped crystal with dimensions of 2.66 $\times$ 0.36 $\times$ 0.28 mm$^3$ was cut precisely along the crystallographic $a$, $b$ and $c$ axes ($b$ is define as the direction perpendicular to the $a$ and $c$ axes \cite{zhao2020realization}) and checked by X-ray diffraction and Laue photograph. All the measurements were carried out using this sample. Resistivity was measured by using the standard four-probe method and thermal conductivity ($\kappa$) was measured by using a conventional steady-state technique with one heater and two thermometers \cite{PhysRevLett.102.167202, PhysRevB.82.094405, PhysRevB.83.174518, PhysRevB.85.134412, PhysRevB.86.174413, PhysRevB.95.224419, PhysRevB.99.104419, PhysRevB.104.104403}. The same four contacts were used for resistivity and thermal conductivity measurements. The electric or heat currents were applied along the $c$ axis, which is the longest dimension, while the magnetic field were applied along either the $a$ or $c$ axis. The transport measurements were carried out in a $^3$He refrigerator equipped with a 14 T magnet. The magnetic field dependencies of resistivity and thermal conductivity at low temperatures were measured after zero-field cooling the sample from 20 K to the target temperatures.

\section{Results and Discussion}

\subsection{Magnetization}

\begin{figure}
\includegraphics[clip,width=8cm]{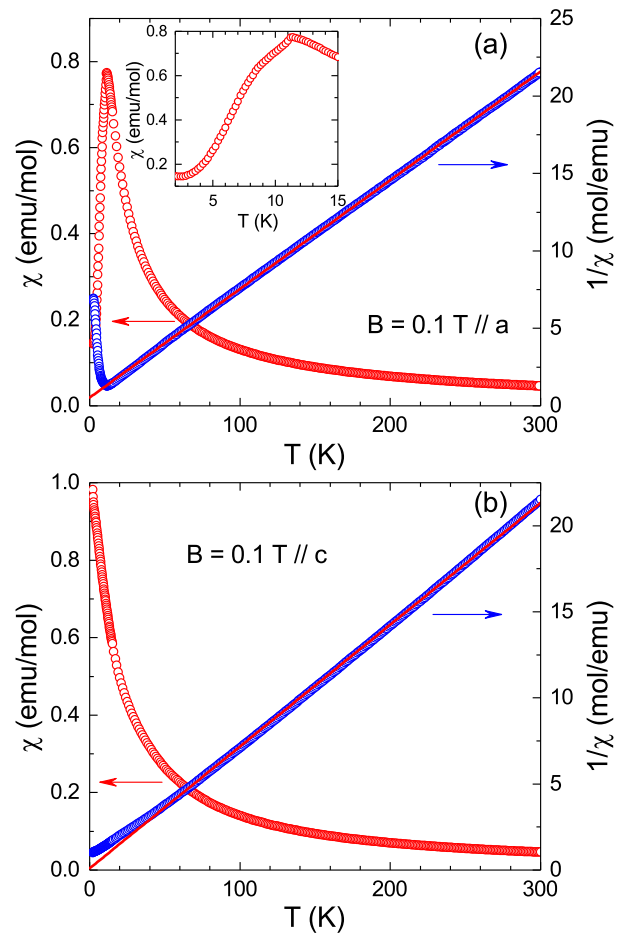}
\caption{Temperature dependence of dc magnetic susceptibility $\chi$ and inverse magnetic susceptibility 1/$\chi$ of HoAgGe single crystal with $B \parallel a$ (a) and $B \parallel c$ (b). The inset shows the low-temperature data of $\chi(T)$. The solid red line indicates the linear fit of the inverse magnetic susceptibility 1/$\chi(T)$ curve.}
\label{M}
\end{figure}

The temperature dependence of magnetic susceptibility $\chi$ measured in 0.1 T field along the $a$ axis is shown in Fig. 1(a). With decreasing temperature, the $\chi(T)$ curve displays a maximum at $\sim$ 11.3 K, corresponding to the antiferromagnetically order, and a weak anomaly at $\sim$ 7 K, corresponding to the spin reorientation transition as previously reported \cite{Morosan2004RAgGe}. In contrast, there is no visible anomaly for the $\chi(T)$ curve with $B \parallel c$. At high temperature ($T >$ 50 K), the 1/$\chi(T)$ curve follows the Curie-Weiss behavior for both $B \parallel a$ and $B \parallel c$. We fit the 1/$\chi(T)$ curve by using the formula $\chi$ = $\chi\rm_0$ + $C$/($T$ - $\theta\rm_{CW}$), where $\chi\rm_0$ is the temperature independent term including the contribution from the core diamagnetism and the Van Vleck paramagnetism, $C$ is the Cuire constant and the parameter $\theta\rm_{CW}$ is the Curie-Weiss temperature. The fitting yields $\theta\rm_{CW}$ = - 7.585 K and $\theta\rm_{CW}$ = - 1.4 K for $B \parallel a$ and $B \parallel c$, respectively. These two values are in agreement with the previous report \cite{Morosan2004RAgGe} and the negative values of $\theta\rm_{CW}$ indicate the predominant AF interactions in HoAgGe.

\begin{figure}
\includegraphics[clip,width=8.5cm]{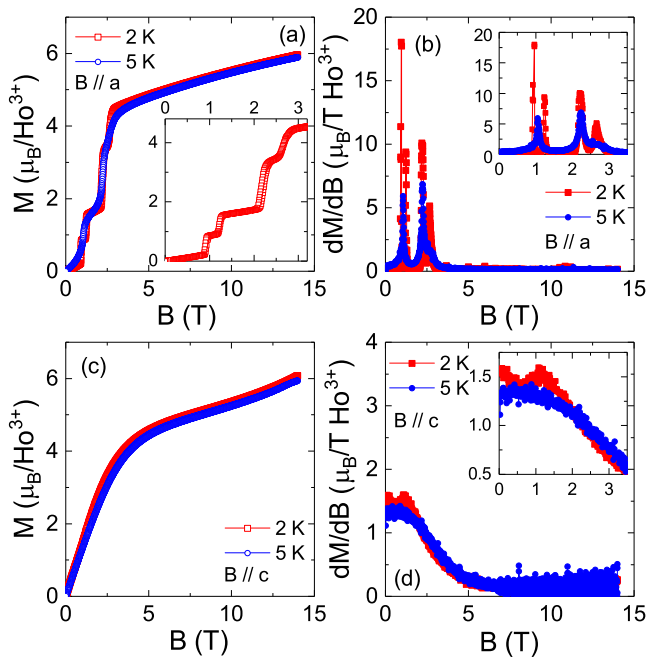}
\caption{Isothermal magnetization of HoAgGe single crystal at $T =$ 2 K and 5 K with $B \parallel a$ (a) or $B \parallel c$ (c). The $dM/dB$ curves for $B \parallel a$ (b) or $B \parallel c$ (d) and the insets are the zoom-in of low field data.}
\label{M}
\end{figure}

Figure 2 shows the magnetization of HoAgGe single crystal with magnetic field along the $a$ or $c$ axis. There are several remarkable features. First, it is clear that the magnetization curves show a series of plateau-like features at low temperatures for $B \parallel a$. At 2 K, the magnetization curve exhibits a narrow 1/6 plateau between 1.0 T and 1.18 T, and the 1/3 plateau between 1.3 T and 2.1 T, as well as another narrow plateau around 2.45 T, probably corresponding to the 7/9 plateau, finally reaches the saturated state above 3 T. According to the previous report, the 1/6 plateau and the 7/9 plateau are only stabilized within a narrow field regime, which is likely related to long wavelength meta-stable structures or order domain walls \cite{zhao2020realization}. At higher temperature of 5 K, the magnetization plateaus become faintness. However, there are three visible metamagnetic transitions at $B \approx$ 1.0, 2.2 and 2.6 T, which are determined by the maxima in the $dM/dB$ curve as shown in Fig. 2(b). Moreover, there is a small hysteresis for $B \parallel a$ at 2 K. In contrast, there is no obvious plateau and hysteresis for magnetization curves with $B \parallel c$, which indicates the strong spin anisotropy in HoAgGe. These experimental results are consistent with the previous study \cite{zhao2020realization}, in which the metamagnetic transitions were explained to result from the competition between the external magnetic field and the weaker next-neighbor couplings. Second, the theoretical value of the saturated magnetization of Ho$^{3+}$ is not reached at 14 T for both field directions, which corresponds to a crystal-field-limited saturated paramagnetic state for $B \parallel a$ and a continuous spin-polarization transition for the upturn of $B \parallel c$ curve, respectively \cite{PhysRevB.71.014445}. Third, the differential d$M$/d$B$ curves for $B \parallel c$ have a small and broad peak at $\sim$ 1 T, which has not been reported in the previous studies. It may indicate some weak metamagnetic transition.

\subsection{Resistivity}

\begin{figure}
\includegraphics[clip,width=8cm]{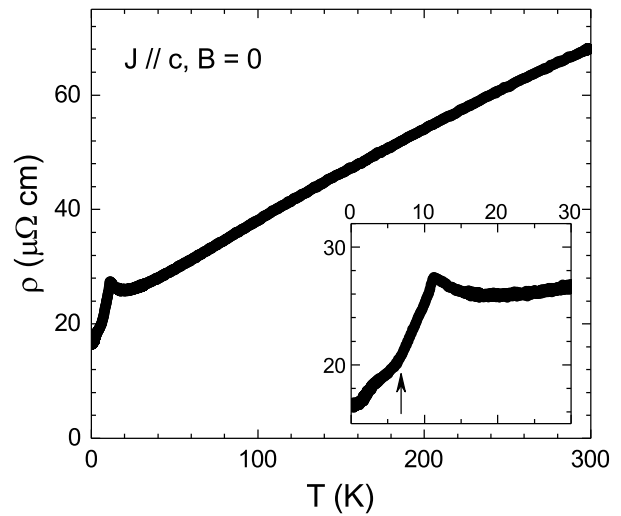}
\caption{Temperature dependence of resistivity for HoAgGe single crystal in zero field. The electric current ($J$) is parallel to the $c$ axis. Inset: zoom-in of the low-temperature data and the arrow indicates the shoulder-like feature around $\sim$ 7 K.}
\label{RT}
\end{figure}

Figure 3 shows the $c$-axis resistivity versus temperature of HoAgGe single crystal, which indicates a good metallic behavior in high temperature range. At low temperatures, there are a sharp peak at $T\rm_N =$ 11.3 K and a shoulder-like feature at $\sim$ 7 K. Note that these data essentially reproduce those in previous reports \cite{zhao2020realization, Morosan2004RAgGe}. It has been proposed that those two features are originated from the AF order and a spin reorientation transition, respectively. Below $T\rm_N$, the loss of spin-disorder scattering, due to the AF order of the Ho$^{3+}$ spins, results in a quick drop in resistivity \cite{Morosan2004RAgGe}.

\begin{figure}
\includegraphics[clip,width=8.5cm]{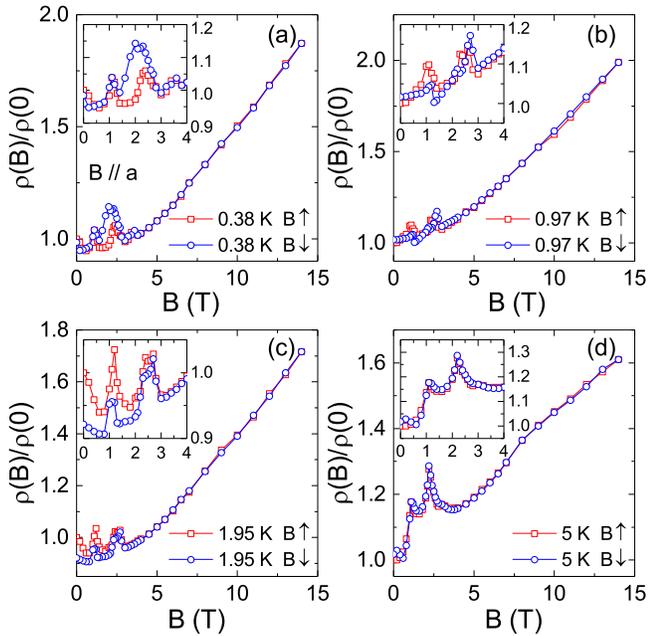}
\caption{Magnetic-field dependence of resistivity for HoAgGe single crystal at 0.38 K (a), 0.97 K (b), 1.95 K (c), and 5 K (d) for $J \parallel c$ and $B \parallel a$. The data shown with red open squares are measured in the field-up process, while the blue open circles show the data in the field-down process. Insert: the zoom-in of $\rho(B)/\rho(0)$ data at 0--4 T, where the hysteresis loops can be seen more clearly.}
\label{RHb}
\end{figure}

To probe the magnetoresistive characteristics of HoAgGe, resistivity as a function of magnetic field was measured with changing field from 0 to 14 T and back to 0 T along the $a$ axis, as shown in Fig. 4. In general, HoAgGe exhibits large and positive transverse MR for $B \parallel a$. In addition, the resistivity displays complex dependencies on the history of applying magnetic field. For example, as shown in Fig. 4(c), at 1.95 K the $\rho(B)/\rho(0)$ curve has three obvious maxima at $B =$ 1.2, 2.4, and 2.7 T, which are close to the transition fields in the magnetization curves; the $\rho(B)/\rho(0)$ monotonically increases with further increasing field ($B >$ 3 T) and the resistivity increases by 60--100 \% at 14 T. Another remarkable feature is that the $\rho(B)/\rho(0)$ curves measured with increasing field are not equal to those with decreasing field at $T \le$ 1.95 K, forming a clear hysteresis at low fields ($B <$ 3 T). The irreversibility weakens gradually with increasing temperature and is not visible at $T =$ 5 K. Apparently, the main electric transport phenomena are closely related to the magnetization behaviors, indicating the spin-electron scatterings.

\begin{figure}
\includegraphics[clip,width=8.5cm]{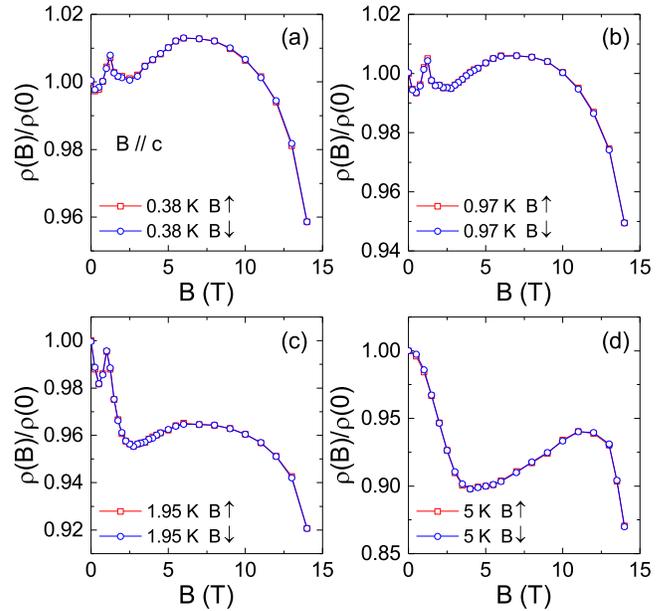}
\caption{Magnetic-field dependence of resistivity for HoAgGe single crystal at 0.38 K (a), 0.97 K (b), 1.95 K (c), and 5 K (d) for $J \parallel c$ and $B \parallel c$. The data shown with red open squares are measured in the field-up process, while the blue open circles show the data in the field-down process.}
\label{RHc}
\end{figure}

Figure 5 shows the $\rho(B)/\rho(0)$ of HoAgGe for $B \parallel c$, which is mainly negative, in contrast to the case of $B \parallel a$. Furthermore, the $\rho(B)/\rho(0)$ curves for the field sweeping-up and sweeping-down along the $c$ axis shows no obvious hysteresis, which is coincided with magnetization measurements. It is obvious that the $\rho(B)/\rho(0)$ curves display similar behavior at all selected temperatures for $B \parallel c$. At $T \le$ 1.95 K, the $\rho(B)/\rho(0)$ curves display a sharp peak at about 1 T which corresponds to the small broad peak in the $dM/dB$ curve for $B \parallel c$ at $T =$ 2 K, followed by a broad peak at high field. The 1 T anomaly of MR data further indicates that there is a weak metamagnetic transition for $B \parallel c$. With increasing temperature, the magnitude of the sharp peak is decreased and disappears at $T =$ 5 K. At high fields ($B >$ 3 T), the $\rho(B)/\rho(0)$ first increases and then drastically decreases, which may be associated with the continuous increase and a small upward of the magnetization curve at high fields.

\subsection{Thermal Conductivity}

\begin{figure}
\includegraphics[clip,width=6.5cm]{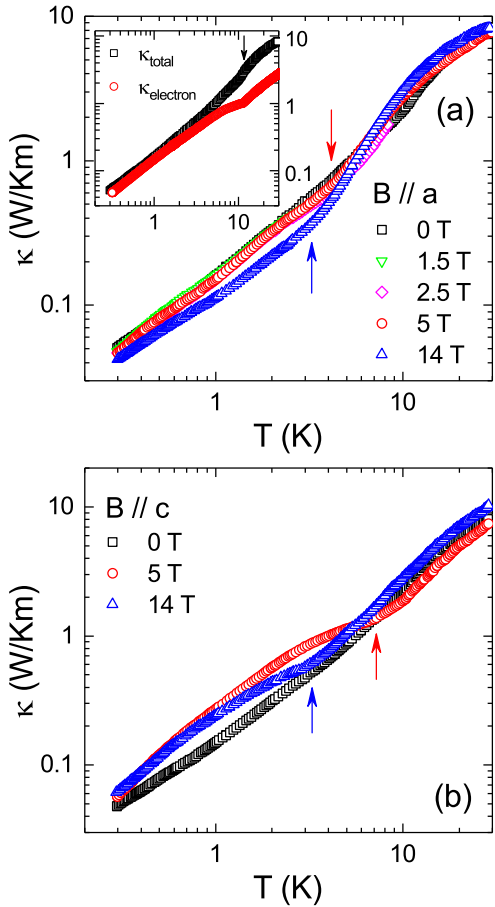}
\caption{Temperature dependence of the $c$-axis thermal conductivity for HoAgGe single crystal at different fields for $B \parallel a$ (a) and $B \parallel c$ (b). The inset shows the temperature dependence of electronic thermal conductivity and the total thermal conductivity at zero field.}
\label{kT}
\end{figure}

Figure 6 shows the temperature dependence of $\kappa$ for HoAgGe single crystal at different magnetic fields along the $a$ and $c$ axis. The zero-field curve displays a weak kink around $T \sim$ 11.7 K, which should be related to the AF transition. With increasing magnetic field along both directions, the kink shifts to lower temperature. Apparently, the direction of magnetic field plays an important role in changing the thermal conductivity. Thermal conductivity of HoAgGe includes phonon and electron contributions. One can make an estimation of electronic thermal conductivity $\kappa\rm_{e}$ with the resistivity data. Assuming Wiedemann-Franz (WF) law is valid and using the formula $\kappa_{e} = LT/\rho$, where $\rho$ is the electrical resistivity and $L$ (= 2.44 $\times$ 10$^{-8}$ W$\Omega$/K$^2$) is the Lorenz number, we can calculate the electronic thermal conductivity. As shown in the inset to Fig. 6(a), below 1 K the calculated $\kappa\rm_{e}$ accounts for 80--95 \% of the total $\kappa$.

\begin{figure}
\includegraphics[clip,width=8.5cm]{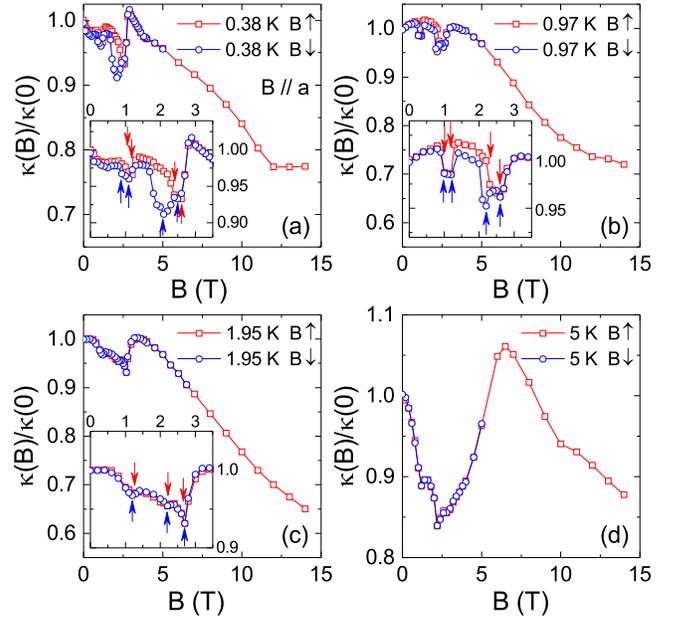}
\caption{Magnetic-field dependence of $c$-axis thermal conductivity for HoAgGe single crystal at 0.38 K (a), 0.97 K (b), 1.95 K (c), and 5 K (d) for $B \parallel a$. The data shown with red open squares are measured in the field-up process, while the blue open circles show the data in the field-down process. Inset: the zoom-in of the low-field data, where the hysteresis loop can be seen more clearly. The position of minima indicated by the arrows.}
\label{kHb}
\end{figure}

Figure 7 shows the magnetic-field dependence of $\kappa$ for HoAgGe single crystal at low temperatures with $B \parallel a$. A notable feature of the $\kappa(B)$ isotherms is that there are some minima at low fields, as indicated by the arrows in Fig. 7, most of which correspond to the maxima of the $\rho(B)/\rho(0)$ curves. It is notable that in many magnetic materials the $\kappa(B)$ exhibits minima at field-induced magnetic transitions \cite{PhysRevB.63.214407, PhysRevLett.98.107201, PhysRevLett.100.137202, PhysRevLett.102.167202, PhysRevB.82.094405, PhysRevB.83.174518, PhysRevB.85.134412, PhysRevB.86.174413}. With further increasing field, the $\kappa(B)/\kappa(0)$ is strongly reduced, which has a good correspondence to the positive MR effect for $B \parallel a$. This clearly indicates that at low temperatures the main heat carriers are electrons. Another phenomenon is that the $\kappa(B)$ isotherms show clear irreversibility in the low field region of 0 $< B <$ 3 T at $T \le$ 1.95 K and the hysteresis loop becomes more pronounced with decreasing temperature, which has good correspondence to the above magnetization and magnetoresistance results.

\begin{figure}
\includegraphics[clip,width=8.5cm]{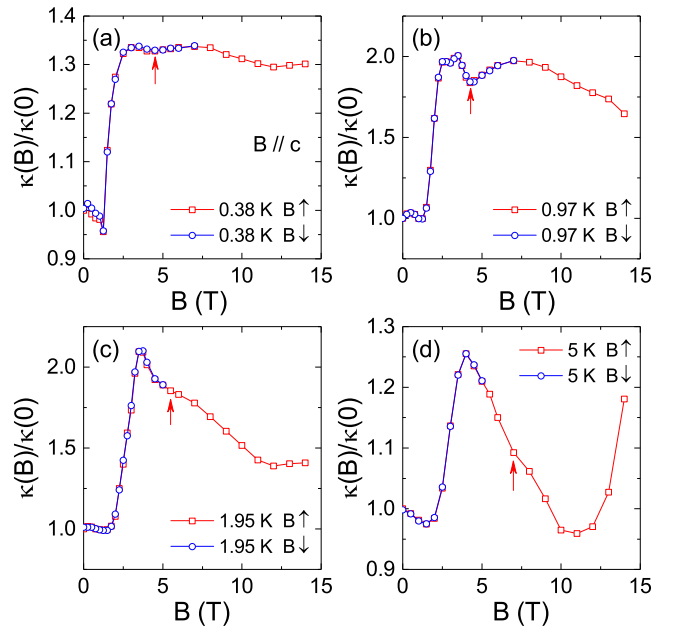}
\caption{Magnetic-field dependence of the $c$-axis thermal conductivity for HoAgGe single crystal at 0.38 K (a), 0.97 K (b), 1.95 K (c), and 5 K (d) for $B \parallel c$. The data shown with red open squares are measured in the field-up process, while the blue open circles show the data in the field-down process. The arrows indicate the position of the anomalies.}
\label{kHc}
\end{figure}

For comparison, the $\kappa(B)/\kappa(0)$ isotherms with $B \parallel c$ are shown in Fig. 8. First, it is notable that the magnitude of magnetothermal conductivity is much larger than that of MR for $B \parallel c$, which indicates that the change of $\kappa$ in this field direction is governed by phonon transport. Second, the $\kappa(B)/\kappa(0)$ curves display some features at low temperatures. There is a minimum at low field, accompanied by a quick increase with increasing field. This low-field minimum has some correspondence to the low-field peak in the MR curves and the magnetization curves with $B \parallel c$, and thus further confirms the unknown metamagnetic transition. In addition, at 0.38 K, there is a small valley around 4.5 T and the valley becomes deeper at $T =$ 0.97 K. With increasing temperature, this valley shifts to higher field and evolves into a slope change, as indicated by the arrows in Fig. 8. Note that there is another broad valley at high field for $T =$ 0.38 and 1.95 K, which has no obvious correspondence to the MR behaviors, so they should be caused by phonons. Finally, at highest temperature of 5 K, the high field behavior is drastically different. There is a broad valley at 11 T, accompanied with strong increase of $\kappa$ at higher fields. Since the spins are already fully polarized at such high magnetic fields, this phenomenon has no relationship to the magnetic excitations. One possibility is the phonon resonant scattering effect by the crystal-field energy levels of rare-earth ions, as many other materials demonstrated \cite{CEF, GBCO, NGSO, EASO}.

\section{Discussions}

First of all, all the magnetization, magnetoresisitance and magnetothermal conductivity results of HoAgGe single crystal indicate a weak metamagnetic transition at $\sim$ 1 T for $B \parallel c$. This phenomenon has not be noticed in previous studies. It would be difficult to understand this transition if the Ho$^{3+}$ spin had an Ising axis in the $ab$ plane \cite{zhao2020realization}. In some earlier works, it was suggested that in the $R$AgGe ($R =$ Tb -- Lu) series the spin anisotropy changes from axial (in TbAgGe) to extreme planar (in TmAgGe) \cite{Morosan2004RAgGe}. Therefore, our present results indicate that the Ho$^{3+}$ spin anisotropy is weaker than Ising type and the spin direction is slightly canted out of the $ab$ plane. This calls for future neutron scattering experiments.

In the case of $B \parallel a$ and $J \parallel c$, HoAgGe displays a positive and large transverse MR up to $\sim$ 100 \% at 14 T and the overall field dependence is between the $B^2$ and $B$-linear behavior, despite of the low-field peaks. It cannot be simply due to the ordinary Lorentz force-induced scattering, which usually induces rather weak MR \cite{Pippard1989}. In general, the two-band theory can give a $B^2$ dependence of MR \cite{Oxford University Press}. In metals with open Fermi Surface, the MR can display $B$-linear behavior \cite{PRSLA1929}. It has been reported for semiconductors that macroscopically disorder leads to the distorted current paths and induces the Hall component, which results in a $B$-linear MR \cite{Nature4262003}. Here, the transverse MR with the field dependence between the $B^2$ and $B$-linear behavior is apparently very unusual. Note that the band structure of HoAgGe is rather complex, with several bands across the Fermi level, and there are both electron and hole carriers \cite{Nature480, Nature486, Nature475}. It is likely that these uncompensate charge carriers are the main reason for the extraordinary transverse MR. Furthermore, the spin scattering of electrons should also play an important role, considering the complex magnetism and magnetic transitions in this material. Indeed, the $\rho(B)/\rho(0)$ curves for both sweeping-up and sweeping-down field show obvious maxima at the metamagnetic transitions; accordingly, the $\kappa(B)/\kappa(0)$ curves show obvious minima at these transitions. These can be attributed to some enhanced electron scattering at the magnetic transitions. Meanwhile, the $\kappa(B)$ with $B \parallel a$ shows a negative magnetothermal effect, which is qualitatively consistent with the positive MR. However, the $\kappa$ is reduced by a factor 20--30 \% at 14 T, as shown in Fig. 7, which is much smaller than the relative change of MR. There are apparently two possible reasons for this. First, the electronic contribution to the total thermal conductivity is much smaller than the phononic one. However, this possibility can be ruled out considering that below 1 K the calculated $\kappa\rm_{e}$ accounts for 80--95 \% of the total $\kappa$. Second, with increasing magnetic field, the phonon scattering by magnetic excitations is significantly suppressed. As a result, the phononic thermal conductivity increases and compensates the decrease of $\kappa\rm_{e}$. This possibility is more likely considering that at 5 K the $\kappa(B)$ is larger than the $\kappa(0)$ for the intermediate fields of 6--8 T.

For $B \parallel c$ and $J \parallel c$, the longitudinal MR is much weaker and mainly negative. For example, the resistivity decreases by only 4--8 \% in field range from 7 to 14 T and at $T \le$ 1.95 K. There are several possible origins of this negative MR. First, it is well known that a negative MR in magnetic materials can be attributed to weakening of spin-dependent scattering with increasing field \cite{JPSJ.45.466, PhysRevB.57.8103, PhysRevB.59.8784, PhysRevB.76.094401, PhysRevLett.113.157202}. Given that the magnetization continuously increases with increasing field, indicating a continuous spin-polarization transition for $B \parallel c$, this origin can be a suitable reason for the negative longitudinal MR in HoAgGe. Second, one possible origin is the current-jetting effect, which is caused by the inhomogeneous spatial distribution of the current in the sample \cite{PhysRevLett.95.186603, Pippard1989}. Since in our measurements, the current was flowing homogenously in the $ac$ plane, this possibility can be excluded. Moreover, the negative MR in HoAgGe does not vanish quickly with raising temperature, which is also inconsistent with the current jetting effect. Third, some topological materials can exhibit some negative MR phenomena due to the chiral anomaly \cite{ong2015science, Liang2015nat.mater, HuiLi10301, PhysRevLett.110.266601}. Considering the complex Fermi surface and high density of states near Fermi level in HoAgGe, the negative MR is unlikely to be caused by chiral anomaly \cite{Nature480, Nature486, Nature475}. In addition, at very low temperatures there is weak positive MR for the intermediate fields. This non-monotonic MR behavior seems to have some correspondence to the magnetothermal conductivity, which displays slight decreasing at low field accompanied with large increasing at high field. However, there are also some clear discrepancy between the MR and the magnetothermal conductivity. That is, the high field enhancement of $\kappa$ is about 100 \% at low temperatures while the negative MR effect is only several percents. This also indicates that the large enhancement of $\kappa$ in high field is mainly due to the suppression of magnetic scattering of phonons. This is similar to the case of $B \parallel a$.

We can also obtain some conclusions on the thermal transport results: (i) at low temperature the phonons are scattered by magnetic excitations, which should be the magnons of the AF state; (ii) magnetic field drives the spin system to the polarized state after going through a series metamagnetic transitions, in which the magnetic scattering of phonons is likely experiencing some sudden changes; (iii) at high fields the phonon scattering by magnetic excitations is strongly suppressed while the resonant scattering by crystal-field levels may be active. It is known that in magnetic materials the thermal conductivity usually exhibits some anomalies at the magnetic transitions, due to either the changes of magnetic excitation transport or spin-phonon scattering \cite{PhysRevB.63.214407, PhysRevLett.98.107201, PhysRevLett.100.137202, PhysRevLett.102.167202, PhysRevB.82.094405, PhysRevB.83.174518, PhysRevB.85.134412, PhysRevB.86.174413, PhysRevB.95.224419, PhysRevB.99.104419, PhysRevB.104.104403}. In particular, the spin ice materials like Dy$_2$Ti$_2$O$_7$, Ho$_2$Ti$_2$O$_7$, and Yb$_2$Ti$_2$O$_7$ have already been found to show anomalies of phonon thermal conductivity at magnetic-field-induced transitions and the peculiar monopole excitations play important role in the heat transport by carrying heat or scattering phonons \cite{PhysRevB.87.144404, PhysRevB.92.094408, PhysRevB.86.060402, PhysRevB.88.054406, PhysRevLett.110.217209, Tokiwa2016}. In HoAgGe, it is unclear whether there is also magnetic monopoles, which are expected elementary excitations for kagome spin ice state. Compared with those insulating AF materials, the heat transport behaviors of intermetallic HoAgGe in magnetic fields are more complicated because of the significant contribution of electron transport. Furthermore, since the electron transport also changes suddenly at the field-induced magnetic transitions, it is difficult for HoAgGe to distinguish how the phonon transport changes at these transitions. Nevertheless, the possible contribution of magnetic monopoles can not be excluded. Apparently, both the electrical and thermal transport properties of HoAgGe deserves further investigations.

\section{SUMMARY}

In summary, the magnetization, resistivity and thermal conductivity of intermetallic HoAgGe single crystals were measured at low temperatures and in magnetic fields along the $a$ and $c$ axis. The magnetization curves display a series of field-induced phase transitions at low temperature for $B \parallel a$, in contrast to the case of $B \parallel c$, which reveals one weak metamagnetic transition. The low-temperature electric and thermal transport along the $c$ axis under magnetic fields along the $a$ or $c$ axis demonstrates some unusual phenomena. In zero field and very low temperatures, the electrons are found to be the main heat carriers for the thermal conductivity. For $B \parallel a$, large and positive MR with extraordinary field dependence is observed and the $\rho(B)$ curves exhibits anomalies at the metamagnetic transitions, accompanied with low-field hysteresis; meanwhile the $\kappa(B)$ mainly decrease with increasing field and display some anomalies at the metamagnetic transitions. All these results have good correspondence to the magnetization behaviors. For $B \parallel c$, weak and negative MR is observed and it is likely due to the weakening of spin-electron scattering upon spin polarization, while the $\kappa(B)$ show rather strong field dependence indicating the role of phonon heat transport.

\begin{acknowledgements}

This work was supported by the National Natural Science Foundation of China (Grant Nos. U1832209, 11874336, 12174361, 12104010, and 12104011) and the Nature Science Foundation of Anhui Province (Grant Nos. 2108085QA22 and 2108085MA16). The work at the University of Tennessee was supported by NSF with Grant No. NSF-DMR-2003117.

\end{acknowledgements}


\begin{thebibliography}{}

\bibitem{NatPhys832}
A. Trebesinger, Nat. Phys. {\bf 4}, 832 (2008).

\bibitem{Nature439}
R. F. Wang, C. Nisoli, R. S. Freitas, J. Li, W. McConville, B. J. Cooley, M. S. Lund, N. Samarth, C. Leighton, V. H. Crespi, and P. Schiffer, Nature {\bf 439}, 303 (2006).

\bibitem{PhysRevB094418}
Y. Qi, T. Brintlinger, and J. Cumings, Phys. Rev. B {\bf 77}, 094418 (2008).

\bibitem{Natphys359}
S. Ladak, D. E. Read, G. K. Perkins, L. F. Cohen, and W. R. Branford, Nat. Phys. {\bf 6}, 359 (2010).

\bibitem{NatPhys68}
E. Mengotti, L. J. Heyderman, A. F. Rodr\'iguez, F. Nolting, R. V. H\"ugli, and H.-B. Braun, Nat. Phys. {\bf 7}, 68 (2011).

\bibitem{Nature553}
S. Zhang, I. Gilbert, C. Nisoli, G.-W. Chern, M. J. Erickson, L. O'Brien, C. Leighton, P. E. Lammert, V. H. Crespi, and P. Schiffer, Nature {\bf 500}, 553 (2013).

\bibitem{RevModPhys1473}
C. Nisoli, R. Moessner, and P. Schiffer, Rev. Mod. Phys. {\bf 85}, 1473 (2013).

\bibitem{NatNanotechnol53}
J. C. Gartside, D. M. Arroo, D. M. Burn, V. L. Bemmer, A. Moskalenko, L. F. Cohen, and W. R. Branford, Nat. Nanotechnol. {\bf 13}, 53 (2018).

\bibitem{JPhysCondensMatter14}
K. Matsuhira, Z. Hiroi, T. Tayama, S. Takagi, and T. Sakakibara, J. Phys.: Condens. Matt. {\bf 14}, 559 (2002).

\bibitem{PhysRevLett257205}
Y. Tabata, H. Kadowaki, K. Matsuhira, Z. Hiroi, N. Aso, E. Ressouche, and B. Fak, Phys. Rev. Lett. {\bf 97}, 257205 (2006).

\bibitem{NatPhys566}
T. Fenell, S. T. Bramwell, D. F. McMorrow, P. Manuel, and A. R. Wildes, Nat. Phys. {\bf 3}, 566 (2007).

\bibitem{PhysRevB.87.144404}
C. Fan, Z. Y. Zhao, H. D. Zhou, X. M. Wang, Q. J. Li, F. B. Zhang, X. Zhao, and X. F. Sun, Phys. Rev. B {\bf 87}, 144404 (2013).

\bibitem{zhao2020realization}
K. Zhao, H. Deng, H. Chen, K. A. Ross, V. Pet\v{r}\'icek, G. G\"unther, M. Russina, V. Hutanu, and P. Gegenwat, Science {\bf 367}, 1218 (2020).

\bibitem{Gibson1996}
R. Gibson, R. P\"ottgen, R. K. Kremer, A. Simon, and K. R. A. Zieback, J. Alloys Compd. {\bf 239}, 34 (1996).

\bibitem{Baran1998}
S. Baran, M. Hofmann, J. Leciejewicz, B. Penc, M. \'Slaski, and A. Szytula, J. Alloys Compd. {\bf 281}, 92 (1998).

\bibitem{Morosan2004RAgGe}
E. Morosan, S. L. Bud'ko, P. C. Canfield, M. S. Torikachvili, and A. H. Lacerda, J. Magn. Magn. Mater. {\bf 277}, 298 (2004).

\bibitem{PhysRevB.71.014445}
E. Morosan, S. L. Bud'ko, and P. C. Canfield, Phys. Rev. B {\bf 71}, 014445 (2005).

\bibitem{PhysRevB.63.214407}
J. Takeya, I. Tsukada, Y. Ando, T. Masuda, K. Uchinokura, I. Tanaka, R. S. Feigelson, and A. Kapitulnik, Phys. Rev. B {\bf 63}, 214407 (2001).

\bibitem{PhysRevLett.98.107201}
A. V. Sologubenko, K. Berggold, T. Lorenz, A. Rosch, E. Shimshoni, M. D. Phillips, and M. M. Turnbull, Phys. Rev. Lett. {\bf 98}, 107201 (2007).

\bibitem{PhysRevLett.100.137202}
A. V. Sologubenko, T. Lorenz, J. A. Mydosh, A. Rosch, K. C. Shortsleeves, and M. M. Turnbull, Phys. Rev. Lett. {\bf 100}, 137202 (2008).

\bibitem{PhysRevLett.102.167202}
X. F. Sun, W. Tao, X. M. Wang, and C. Fan, Phys. Rev. Lett. {\bf 102}, 167202 (2009).

\bibitem{PhysRevB.82.094405}
X. M. Wang, C. Fan, Z. Y. Zhao, W. Tao, X. G. Liu, W. P. Ke, X. Zhao, and X. F. Sun, Phys. Rev. B {\bf 82}, 094405 (2010).

\bibitem{PhysRevB.83.174518}
Z. Y. Zhao, X. M. Wang, B. Ni, Q. J. Li, C. Fan, W. P. Ke, W. Tao, L. M. Chen, X. Zhao, and X. F. Sun, Phys. Rev. B {\bf 83}, 174518 (2011).

\bibitem{PhysRevB.85.134412}
Z. Y. Zhao, X. G. Liu, Z. Z. He, X. M. Wang, C. Fan, W. P. Ke, Q. J. Li, L. M. Chen, X. Zhao, and X. F. Sun, Phys. Rev. B {\bf 85}, 134412 (2012).

\bibitem{PhysRevB.86.174413}
X. M. Wang, Z. Y. Zhao, C. Fan, X. G. Liu, Q. J. Li, F. B. Zhang, L. M. Chen, X. Zhao, and X. F. Sun, Phys. Rev. B {\bf 86}, 174413 (2012).

\bibitem{PhysRevB.95.224419}
J. D. Song, X. M. Wang, Z. Y. Zhao, J. C. Wu, J. Y. Zhao, X. G. Liu, X. Zhao, and X. F. Sun, Phys. Rev. B {\bf 95}, 224419 (2017).

\bibitem{PhysRevB.99.104419}
X. Zhao, Z. Y. Zhao, L. M. Chen, X. Rao, H. L. Che, L. G. Chu, H. D. Zhou, L. S. Ling, J. F. Wang, and X. F. Sun, Phys. Rev. B {\bf 99}, 104419 (2019).

\bibitem{PhysRevB.104.104403}
N. Li, Q. Huang, A. Brassington, X. Y. Yue, W. J. Chu, S. K. Guang, X. H. Zhou, P. Gao, E. X. Feng, H. B. Cao, E. S. Choi, Y. Sun, Q. J. Li, X. Zhao, H. D. Zhou, and X. F. Sun, Phys. Rev. B {\bf 104}, 104403 (2021).

\bibitem{CEF}
A. Rassili, K. Durczewski, and M. Ausloos, Phys. Rev. B {\bf 58}, 5665 (1998).

\bibitem{GBCO}
X. F. Sun, A. A. Taskin, X. Zhao, A. N. Lavrov, and Y. Ando, Phys. Rev. B {\bf 77}, 054436 (2008).

\bibitem{NGSO}
Q. J. Li, Z. Y. Zhao, H. D. Zhou, W. P. Ke, X. M.Wang, C. Fan, X. G. Liu, L. M. Chen, X. Zhao, and X. F. Sun, Phys. Rev. B {\bf 85}, 174438 (2012).

\bibitem{EASO}
H. L. Che, Z. Y. Zhao, X. Rao, L. G. Chu, N. Li, W. J. Chu, P. Gao, X. Y. Yue, Y. Zhou, Q. J. Li, Q. Huang, E. S. Choi, Y. Y. Han, Z. Z. He, H. D. Zhou, X. Zhao, and X. F. Sun, Phys. Rev. Materials {\bf 4}, 054406 (2020).

\bibitem{Pippard1989}
A. B. Pippard, {\it Magnetoresistance in Metals} (Cambridge University Press, 1989).

\bibitem{Oxford University Press}
J. M. Ziman, {\it Electrons and Phonons: Theory of Transport Phenomena in Solids} (Oxford Unibersity Press, 1960).

\bibitem{PRSLA1929}
P. Kapitza, Proc. R. Soc. Lond. A {\bf 123}, 292 (1929).

\bibitem{Nature4262003}
M. M. Parish and P. B. Littlewood, Nature {\bf 426}, 162 (2003).

\bibitem{Nature480}
M. G. Vergniory, L. Elcoro, C. Felser, N. Regnault, B. A. Bernevig, and Z. Wang, Nature {\bf 566}, 480 (2019).

\bibitem{Nature486}
F. Tang, H. C. Po, A. Vishwanath, and X. Wan, Nature {\bf 566}, 486 (2019).

\bibitem{Nature475}
T. Zhang, Y. Jiang, Z. Song, H. Huang, Y. He, Z. Fang, H. Wang, and C. Fang, Nature {\bf 566}, 475 (2019).

\bibitem{JPSJ.45.466}
K. Usami, J. Phys. Soc. Jpn. {\bf 45}, 466 (1978).

\bibitem{PhysRevB.57.8103}
J. Y. Chan, S. M. Kauzlarich, P. Klavins, R. N. Shelton, and D. J. Webb, Phys. Rev. B {\bf 57}, 8103(R) (1998).

\bibitem{PhysRevB.59.8784}
Z. Zeng, M. Greenblatt, and M. Croft, Phys. Rev. B {\bf 59}, 8784 (1999).

\bibitem{PhysRevB.76.094401}
J. Du, D. Li, Y. B. Li, N. K. Sun, J. Li, and Z. D. Zhang, Phys. Rev. B {\bf 76}, 094401 (2007).

\bibitem{PhysRevLett.113.157202}
Y. Guo, J. Dai, J. Zhao, C. Wu, D. Li, L. Zhang, W. Ning, M. Tian, X. C. Zeng, and Y. Xie, Phys. Rev. Lett. {\bf 113}, 157202 (2014).

\bibitem{PhysRevLett.95.186603}
J. Hu, T. F. Rosebaum, and J. B. Betts, Phys. Rev. Lett. {\bf 95}, 186603 (2005).

\bibitem{ong2015science}
J. Xiong, S. K. Kushwaha, T. Liang, J. W. Krizan, M. Hirschberger, W. Wang, R. J. Cava, and N. P. Ong, Science {\bf 350}, 413 (2015).

\bibitem{Liang2015nat.mater}
T. Liang, Q. Gibson, M. N. Ali, M. Liu, R. J. Cava, and N. P. Ong, Nat. Mater. {\bf 14}, 280 (2015).

\bibitem{HuiLi10301}
H. Liu, H. He, H. Z. Lu, H. Zhang, H. Liu, R. Ma, Z. Fan, S. Q. Shen, and J. Wang, Nat. Commun. {\bf 7}, 10301 (2016).

\bibitem{PhysRevLett.110.266601}
B. Fauqu\'e, D. LeBoeuf, B. Vignolle, M. Nardone, C. Proust, and K. Behnia, Phys. Rev. Lett. {\bf 110}, 266601 (2013).

\bibitem{PhysRevB.92.094408}
S. J. Li, Z. Y. Zhao, C. Fan, B. Tong, F. B. Zhang, J. Shi, J. C. Wu, X. G. Liu, H. D. Zhou, X. Zhao, and X. F. Sun, Phys. Rev. B {\bf 92}, 094408 (2015).

\bibitem{PhysRevB.86.060402}
G. Kolland, O. Breunig, M. Valldor, M. Hiertz, J. Frielingsdorf, and T. Lorenz, Phys. Rev. B {\bf 86}, 060402(R) (2012).

\bibitem{PhysRevB.88.054406}
G. Kolland, M. Valldor, M. Hiertz, J. Frielingsdorf, and T. Lorenz, Phys. Rev. B {\bf 88}, 054406 (2013).

\bibitem{PhysRevLett.110.217209}
W. H. Toews, S. S. Zhang, K. A. Ross, H. A. Dabkowska, B. D. Gaulin, and R. W. Hill, Phys. Rev. Lett. {\bf 110}, 217209 (2013).

\bibitem{Tokiwa2016}
Y. Tokiwa, T. Yamashita1, M. Udagawa, S. Kittaka, T. Sakakibara, D. Terazawa, Y. Shimoyama, T. Terashima, Y. Yasui, T. Shibauchi, and Y. Matsuda, Nat. Commun. {\bf 7}, 10807 (2016).

\end{thebibliography}
\end{document}